\documentclass[a4paper,11pt]{article}
\usepackage{jcappub}
\usepackage{enumerate}
\usepackage{booktabs}
\usepackage{bm}
\usepackage{xcolor}
\usepackage{subcaption}
\usepackage{multirow}
\arxivnumber{2511.16256}

\title{\boldmath Accelerating reionization constraints: An ANN-emulator framework for the SCRIPT Semi-numerical Model}

\author[1]{Saptarshi Sarkar\note{Corresponding author.}}
\author{and Tirthankar Roy Choudhury}
\affiliation{National Centre for Radio Astrophysics, Tata Institute of Fundamental Research,\\
Pune University Campus, Ganeshkhind, Pune 411007, India}

\emailAdd{sarkar@ncra.tifr.res.in}
\emailAdd{tirth@ncra.tifr.res.in}

\abstract{
Constraining the Epoch of Reionization (EoR) with physically motivated simulations is hampered by the high cost of conventional parameter inference. We present an efficient emulator-based framework that dramatically reduces this bottleneck for the photon-conserving semi-numerical code \texttt{SCRIPT}. Our approach combines (i) a reliable coarse-resolution MCMC to locate the high-likelihood region (exploiting the large-scale convergence of \texttt{SCRIPT}) with (ii) an adaptive, targeted sampling strategy to build a compact high-resolution training set for an artificial neural network based emulator of the model likelihood. With only $\approx 10^3$ high-resolution simulations, the trained emulators achieve excellent predictive accuracy ($R^2 \approx 0.97$--$0.99$) and, when embedded within an MCMC framework, reproduce posterior distributions from full high-resolution runs. Compared to conventional MCMC, our pipeline reduces the number of expensive simulations by a factor of $\sim 100$ and lowers total CPU cost by up to a factor of $\sim 70$, while retaining statistical fidelity. This computational speedup makes inference in much higher-dimensional models tractable (e.g., those needed to incorporate JWST and upcoming 21\,cm datasets) and provides a general strategy for building efficient emulators for next generation of EoR constraints.
}


\begin{document}
\maketitle
\flushbottom

\section{Introduction} \label{sec:intro}

The Epoch of Reionization (EoR) marks the last major phase transition in the universe,  during which the intergalactic medium (IGM) was transformed from a cold, neutral state into a hot, ionized one~\cite{barkanaBeginningFirstSources2001}. Despite substantial theoretical and observational progress, the detailed timeline and physical drivers of reionization remain uncertain.  A new generation of high-redshift observations, from facilities such as the Low-Frequency Array (LOFAR)~\cite{haarlemLOFARLOwFrequencyARray2013}, Murchison Widefield Array (MWA)~\cite{tingayMurchisonWidefieldArray2013}, Square Kilometre Array (SKA)~\cite{koopmansCosmicDawnEpoch2015}, Hydrogen Epoch of Reionization Array (HERA)~\cite{deboerHydrogenEpochReionization2017}, Atacama Large Millimeter/submillimeter Array (ALMA) \cite{woottenAtacamaLargeMillimeter2009}, and the James Webb Space Telescope (JWST)~\cite{gardnerJamesWebbSpace2023}, is already delivering, or will soon deliver, unprecedented constraints on the underlying astrophysical processes. Fully exploiting these data, however, requires efficient parameter inference based on numerical or semi-numerical simulations capable of modeling the coupled evolution of the ionization state and thermal history of the IGM, together with galaxy populations.

Traditional approaches rely on Markov chain Monte Carlo (MCMC) sampling, repeatedly evaluating simulations across multi-dimensional parameter spaces. While robust, this strategy quickly becomes computationally prohibitive: even semi-numerical reionization models require tens of thousands of simulation evaluations for convergence, making inference increasingly infeasible as model complexity grows.

In recent years, a variety of machine learning (ML) techniques have been proposed to accelerate parameter inference in cosmological studies. These methods have been applied to infer parameters directly from the 21\,cm power spectrum~\cite[e.g.,][]{shimabukuroAnalysing21Cm2017, doussotImprovedSupervisedLearning2019, choudhuryExtracting21cmPower2022, shimabukuroEstimationIiBubble2022, choudhuryInferringIGMParameters2025}, global signal~\cite[e.g.,][]{choudhuryExtracting21Cm2020,choudhuryUsingArtificialNeural2021}, as well as from images~\cite[e.g.,][]{gilletDeepLearning21cm2019, kwonDeepLearningStudy21cm2020, neutschInferringAstrophysicsDark2022}, and lightcones~\cite[e.g.,][]{zhaoSimulationbasedInferenceReionization2022, prelogovicMachineLearningAstrophysics2022}. In particular, ML-based emulators have emerged as powerful surrogates for computationally expensive simulations. Trained on a representative set of simulations (the training dataset), these emulators can efficiently interpolate simulation outputs across high-dimensional parameter spaces with remarkable accuracy, offering results at a fraction of the computational cost of traditional methods. Among the various ML approaches used to construct emulators, artificial neural networks (ANNs) have been particularly popular~\cite[e.g.,][]{schmitEmulationReionizationSimulations2018, jenningsEvaluatingMachineLearning2019, cohenEmulatingGlobal21cm2020,
mondalTightConstraintsExcess2020, bevinsGlobalemuNovelRobust2021, bye21cmVAEVeryAccurate2022, tiwariImprovingConstraintsReionization2022, breitman21cmemuEmulator21cmfast2023, sikderEmulationCosmicDawn2024, lazareHERABoundXray2024, meriotLORELIDatabase212024, choudhuryInferringIGMParameters2025, mahidaANNBNNInferring2025, maityEmulatorbasedForecastingAstrophysics2025, gonzalez-hernandezUsingNeuralEmulators2025}, while Gaussian processes (GPs) have also seen widespread use~\cite[e.g.,][]{kernEmulatingSimulationsCosmic2017, maityFastMethodReionization2023, choudhuryGPRbasedEmulatorSeminumerical2024, gharaConstrainingIntergalacticMedium2020}. More recently, advanced deep learning architectures have been explored, including convolutional autoencoders and vision transformer-based models for emulating reionization maps~\cite[e.g.,][]{chardinDeepLearningModel2019, postureCosmoUiTVisionTransformerUNet2025}, as well as generative adversarial networks (GANs) for lightcone image generation~\cite[e.g.,][]{diaoMultifidelityEmulatorLargescale2025}.

However, a central challenge in emulator construction lies in generating an effective training dataset. In practice these training datasets are generated using outputs of simulations evaluated at parameter points sampled from the model's parameter space. Common sampling strategies include simple approaches, such as uniform random and grid sampling~\cite[e.g.,][]{chardinDeepLearningModel2019, cohenEmulatingGlobal21cm2020, sikderEmulationCosmicDawn2024, meriotLORELIDatabase212024, choudhuryInferringIGMParameters2025, facchinettiNeuralNetworkEmulation2025, mahidaANNBNNInferring2025, gharaConstrainingIntergalacticMedium2020, maityEmulatorbasedForecastingAstrophysics2025, postureCosmoUiTVisionTransformerUNet2025}. Uniform random sampling is easy to implement and ensures no bias towards any subregion, but it can lead to uneven coverage: samples can cluster by chance or leave gaps, especially in high dimensions. This strategy is effective if one can afford a very large sample size and the emulated quantity varies fairly smoothly with the parameters. However, it becomes inefficient in high-dimensional parameter spaces, as many samples are required to adequately cover the space, increasing the chances of large unsampled regions and redundant sampling along certain directions. By contrast, a grid sampling approach places points at fixed intervals along each parameter axis avoiding the problem of uneven coverage. But, this approach also suffers from a severe drawback: because each parameter is held fixed on a lattice, many samples end up sharing the same value in one or more parameters. This wastes computation time and means slowly-varying directions are over-sampled. Moreover, the total number of grid points required to cover the space grows exponentially with the dimensionality, leading to poor scalability to high-dimensional parameter spaces.

A widely used alternative strategy is Latin hypercube (LH) sampling~\cite[e.g.,][]{schmitEmulationReionizationSimulations2018, jenningsEvaluatingMachineLearning2019, tiwariImprovingConstraintsReionization2022, lazareHERABoundXray2024, cooperSimulationBasedInference2025, gonzalez-hernandezUsingNeuralEmulators2025, diaoMultifidelityEmulatorLargescale2025}. In LH sampling, each parameter's range is divided into $N$ equal-probability intervals, and one sample is drawn in each interval for each parameter. The $N$ samples are arranged so that no two share the same value in any parameter. This ensures that each parameter is evenly covered and avoids duplicate marginal values. Crucially, this sampling strategy overcomes the scalability limitations of uniform random and grid sampling methods. While this strategy performs well when parameters are reasonably constrained and the priors are narrow, it becomes inefficient when parameters are poorly constrained and the priors are broad. In such cases, the emulated quantity may vary rapidly across the parameter space, requiring a large number of samples to adequately capture these variations. However, many of these samples will correspond to low-likelihood regions when used for inference, leading to significant computational waste and slower emulator development despite the increased sampling effort.

A possible solution to the above problem was demonstrated in Kern et al.~\cite{kernEmulatingSimulationsCosmic2017}, where they first employed coarse LH samples spanning a broad parameter range to build an approximate emulator and performed a preliminary MCMC exploration to locate the high-likelihood region of the parameter space. They then refined the emulator by adding denser, spherical training sets around the estimated maximum a posteriori point. A related idea appears in Breitman et al.~\cite{breitman21cmemuEmulator21cmfast2023}, where they trained the emulator using samples drawn from a previous inference run, ensuring dense coverage of the high-likelihood region in parameter space. Although both strategies mitigate the inefficiency of sampling over broad priors by concentrating computational effort within regions of high-likelihood, they still have their drawbacks. In the first approach, a preliminary MCMC exploration using an emulator trained on coarse LH samples drawn from a broad prior can lead to spurious posterior distributions that have little to no overlap with the true posterior obtained from a full MCMC run. Consequently, the inferred high-likelihood region may be systematically misplaced, greatly reducing the reliability of subsequent refinements. In the second approach, while utilizing samples from previous inference runs ensures dense coverage of the high-likelihood region, it relies on the availability of existing posterior samples that accurately represent the model of interest. Such data may not exist for new or modified models, or for simulations with expanded parameter spaces, thereby limiting the general applicability of this method.

In this work, we introduce an ANN-based likelihood emulator for the semi-numerical code \texttt{SCRIPT}~\cite{choudhuryPhotonNumberConservation2018, maityProbingThermalHistory2022}. This framework is designed to overcome the challenges associated with existing targeted sampling methods for training dataset generation by exploiting a key property of \texttt{SCRIPT}: its numerical convergence at coarse resolutions for large-scale properties relevant to observational constraints~\cite{maityFastMethodReionization2023}. This key property allows us to build a framework that directly overcomes the challenges of existing targeted sampling methods. First, to solve the potential unreliability of preliminary MCMC explorations built on coarse, non-convergent samples~\cite{kernEmulatingSimulationsCosmic2017}, we leverage the convergence of \texttt{SCRIPT} to run a full, reliable MCMC using inexpensive coarse-resolution simulations. This ensures our training samples are drawn directly from the true high-likelihood region, as developed and implemented in previous works~\cite{maityFastMethodReionization2023,choudhuryGPRbasedEmulatorSeminumerical2024}. Second, to remain general-purpose and not rely on pre-existing inference runs~\cite{breitman21cmemuEmulator21cmfast2023}, our method generates this training set from scratch. 

To further optimize sampling efficiency, we introduce an adaptive training-size procedure that iteratively increases the number of high-resolution simulations used for training until the resulting emulators converge according to a Kullback-Leibler (KL) divergence criterion. This ensures that the number of expensive high-resolution simulations, typically the dominant computational cost, is kept to the minimum required for reliable performance. Using this framework, we develop ANN-based emulators for the likelihood for a five-parameter \texttt{SCRIPT} model that includes inhomogeneous recombinations, spatially varying thermal history, and radiative feedback. Embedded within a conventional MCMC, these emulators reproduce the full high-resolution posterior distributions with excellent fidelity while reducing the number of expensive simulation evaluations, and thus establishes a scalable, robust, and general-purpose workflow for emulator-based inference in reionization studies. 

The remainder of this paper is organized as follows. In section~\ref{sec:conventional_mcmc}, we describe the reionization model, outline the observational constraints, and present the results of a conventional MCMC analysis that serve as a benchmark. In section~\ref{sec:ann_emulator}, we first provide a brief overview of ANNs, then describe generation of the training dataset and the training procedure, and finally compare the parameter constraints obtained from conventional MCMC with those from ANN-based emulators. We summarize our key findings in section~\ref{sec:summary}. The cosmological parameters used in this work are $\Omega_m = 0.308, \, \Omega_\Lambda=1-\Omega_m, \,\Omega_b=0.0482, \,h=0.678, \,\sigma_8=0.829,\text{ and }n_s=0.961$~\cite{adePlanck2015Results2016}.

\section{Conventional MCMC analysis} \label{sec:conventional_mcmc}

In this section, we describe the reionization model, observational constraints, and results of conventional MCMC-based parameter inference analyses. These will serve as the benchmark for comparing our emulator-based parameter constraints in the next section. This section draws closely on previous works~\cite{maityConstrainingReionizationThermal2022, maityFastMethodReionization2023}, and is included for completeness.

\subsection{The reionization model} \label{sec:model}

The model used in this work is based on the semi-numerical photon conserving framework \texttt{SCRIPT}~\cite{choudhuryPhotonNumberConservation2018}, which provides the ionization state of the universe in a cosmologically representative simulation volume. The main feature of this framework is its ability to explicitly conserve the ionizing photons and hence produce numerically convergent power spectra of ionization fluctuations with respect to the resolution of the ionization maps. In particular, we use a five-parameter \texttt{SCRIPT} model that is extended to incorporate inhomogeneous recombinations and to compute the thermal history, thereby enabling the self-consistent inclusion of radiative feedback effects~\cite{maityProbingThermalHistory2022, maityConstrainingReionizationThermal2022, maityFastMethodReionization2023}. For a given redshift, we supply our model with the density field and the distribution of collapsed halos capable of producing ionizing radiation. The model then produces the corresponding ionization field at that redshift. To generate the density fields, we use the second-order Lagrangian perturbation theory (2LPT) approximation rather than a full N-body simulation, as our focus is solely on capturing the large-scale features of the intergalactic medium (IGM). In particular, we use \texttt{GADGET-2}~\cite{springelCosmologicalSimulationCode2005a} plugins provided by the 2LPT density field generator \texttt{MUSIC}~\cite{hahnMultiscaleInitialConditions2011} to generate the density fields. For computing the distribution of collapsed halos, we use a sub-grid prescription based on the conditional ellipsoidal mass function~\cite{shethExcursionSetModel2002}. In this work, we use a simulation box size of $256~h^{-1}$~cMpc, which is large enough for the observables that we are interested in computing. We use simulation boxes with redshifts from $z=5$ to $20$ with an interval of $\Delta z = 0.1$, which enables us to compute the reionization history.

Our model requires the ionization efficiency $\zeta$, which is used to compute the ionization maps by estimating the number of available ionizing photons per hydrogen atom. To include inhomogeneous recombinations, we tune the ionization criteria to compensate for excess neutral atoms, and incorporate small-scale fluctuations through a globally averaged clumping factor $C_{\mathrm{HII}}$, fixing its value $C_{\mathrm{HII}}=3$ motivated from earlier simulation studies~\cite{daloisioHydrodynamicResponseIntergalactic2020}. The model also solves for the thermal history of each grid cell in the box and accounts for effects of spatially inhomogeneous reionization on the thermal evolution by assuming that a region's temperature increments by a value $T_{\mathrm{re}}$, known as the reionization temperature, as it is ionized for the first time. In addition, our model also incorporates radiative feedback by suppressing production of ionizing photons in halos where the gas is heated up. Several methods of incorporating this feedback have been introduced in~\cite{maityProbingThermalHistory2022}. In this work, we use the ``step feedback'' model.

In addition to the ionization history, we compute the galaxy UV luminosity function (UVLF) by relating the ionizing properties of the sources of reionization to their non-ionizing UV emission. The ionization efficiency $\zeta$ is a multiplicative combination of the star formation efficiency $f_\star$ and the escape fraction of ionizing photons $f_{\mathrm{esc}}$. While the UVLF is sensitive only to $f_\star$, computing it from $\zeta$ requires independent knowledge of $f_{\mathrm{esc}}$ (or equivalently $f_\star$) in order to disentangle these two quantities. An additional ingredient required to compute the UVLF is the relation between the ionizing and non-ionizing UV properties of the stellar population. We specify this using the slope of the galaxy spectrum above the Lyman limit, $\alpha_s$, and the ratio of luminosities at the Lyman continuum (corresponding to rest wavelength 912~\AA) and at 1500~\AA, $\mathcal{R}_{912/\mathrm{UV}} \equiv L_{912}/L_{\mathrm{UV}}$. As in previous works \cite{maityProbingThermalHistory2022,maityConstrainingReionizationThermal2022}, we adopt $\alpha_s = 2$ and $\mathcal{R}_{912/\mathrm{UV}} = 0.2$, consistent with galaxies having a Salpeter IMF and metallicity $Z \approx 0.1\,Z_\odot=0.002$, as estimated using \texttt{Starburst99} \cite{leithererStarburst99SynthesisModels1999}. We note that both these stellar population parameters are degenerate with the escape fraction. We also note that the UV continuum is modulated by dust attenuation \cite[e.g.,][]{chisholmAccuratelyPredictingEscape2018,gazagnesOriginEscapeLyman2020,maNoMissingPhotons2020}. However, because this effect is largely degenerate with the star formation efficiency $f_\star$, we do not attempt to model dust attenuation separately in our analysis.

Within this framework, our model is described by five free parameters as described below:

\begin{enumerate}
    \item We assume the ionization efficiency to be independent of the halo mass $M_{\mathrm{h}}$, and to have a power-law redshift dependence of the form
    \begin{equation}
        \zeta(z)=\zeta_0\left(\frac{10}{1+z}\right)^\alpha,
    \end{equation}
    where $\zeta_0$ is defined as the ionization efficiency at $z=9$, and $\alpha$ is the power-law index. We treat $\log\zeta_0$ and $\alpha$ as free parameters in our model.
    \item The reionization temperature $T_{\mathrm{re}}$ is used by our model to compute the thermal evolution of the IGM. We treat $\log T_{\mathrm{re}}$ as a free parameter in our model.
    \item Finally, as described above, the model needs the escape fraction $f_{\mathrm{esc}}$, which is the fraction of photons that escape from halos of mass $M_h$, for calculating the UVLFs. Since we only use UVLF data from $z=6$ and $7$ as observational constraints, we ignore its $z$ dependence and parameterize the escape fraction as
    \begin{equation}
        f_{\mathrm{esc}}(M_h)=f_{\mathrm{esc},0}\left(\frac{M_h}{10^{9}M_\odot}\right)^\beta
    \end{equation}
    where $f_{\mathrm{esc},0}$ is defined as the escape fraction at $M_h=10^{9}M_\odot$, and $\beta$ is the power-law index.
    We treat $\log f_{\mathrm{esc},0}$ and $\beta$ as free parameters in our model.

    Because $\zeta \propto f_\star f_{\mathrm{esc}}$, assuming a halo mass independent ionization efficiency $\zeta$ while introducing a halo mass dependence in the escape fraction of the form $f_{\mathrm{esc}} \propto M_h^{\beta}$ implies an effective inverse halo mass dependence for the star formation efficiency, $f_\star \propto M_h^{-\beta}$. We emphasize that this should be interpreted as an effective scaling arising from the adopted parametrization, rather than as an assumption that the star formation efficiency is intrinsically independent of halo mass. Indeed, numerous studies have shown that feedback-regulated star formation leads to a strong halo mass dependence of $f_\star$, particularly at $z \sim 6$--$7$, which plays a key role in shaping the UVLF \cite[e.g.,][]{sunConstraintsStarFormation2016,tacchellaDustAttenuationBulge2018,sippleStarFormationEfficiency2024}. Within this effective framework, the resulting scaling of $f_\star$ is broadly consistent with parameter estimates obtained in earlier studies \cite[e.g.,][]{parkInferringAstrophysicsReionization2019,qinReionizationGalaxyInference2021}. However, we note that this assumption is adopted here as a simplifying choice. In the future, as additional observational data become available, we plan to constrain an extended version of our model~\cite{choudhuryCapturingSmallScaleReionization2025}, in which both $f_{\mathrm{esc}}$ and $f_\star$ are allowed to have independent halo mass dependencies.
    
\end{enumerate}
To summarize, our model has the following five free parameters: $\{\log\zeta_0,\,\alpha,\,\log T_{\mathrm{re}},\,\log f_{\mathrm{esc},0},\,\beta\}$.

\subsection{Observational constraints} \label{sec:observations}

The observational constraints used in this work are similar to those used in our previous works~\cite{maityConstrainingReionizationThermal2022, maityFastMethodReionization2023}. These are summarized below:

\begin{enumerate}
    
    \item We use the latest Planck measurement of the CMB optical depth $\tau_e = 0.054\pm 0.007$~\cite{aghanimPlanck2018Results2020}.
    
    \item We employ model-independent lower bounds on the ionized fraction derived from the dark-pixel analysis of quasar spectra by McGreer et al.~\cite{mcgreerModelindependentEvidenceFavour2015}. These measurements provide lower limits on the ionization fraction at $z=5.6$ and $z=5.9$. \label{constraint:darkpixels} More recent analyses~\cite{jinNearlyModelindependentConstraints2023,zhuProbingUltralateReionization2023,daviesUpdatedDarkPixel2025} have updated these constraints using improved datasets, in some cases tightening the limits and in others relaxing them. For methodological continuity and comparability with our earlier work, however, we adopt the values reported in ref.~\cite{mcgreerModelindependentEvidenceFavour2015}.
    
    \item We assume that reionization is complete by $z = 5$. This requirement is less restrictive than current constraints from Ly$\alpha$ absorption spectra, which indicate that reionization likely concluded by $z \approx 5.3-5.6$~\cite[e.g.,][]{bosmanHydrogenReionizationEnds2022}. A further motivation for this choice is practical: our simulations extend only down to $z=5$, and when computing the CMB optical depth $\tau_e$ we assume the universe is fully ionized at lower redshifts. We therefore discard all reionization histories that do not reach full ionization by $z=5$. To avoid a sharp cutoff in the likelihood, which can lead to instabilities during ANN training, we impose this condition as a smooth lower bound on the ionized fraction at $z=5$, implemented via an error-function window of width 0.01. \label{constraint:reionzend}
    
    \item We utilize data on galaxy UVLFs at $z=6$ and $7$ obtained from optical studies~\cite{bouwensUVLUMINOSITYFUNCTIONS2015,bouwens6LuminosityFunction2017}. We do not incorporate the latest JWST UVLF measurements in this work. The main goal of this work is to develop and validate an ANN-emulator framework for parameter inference and to benchmark its performance against that of a conventional MCMC. Accurately modeling the evolving galaxy UVLFs over the full redshift range $6 \lesssim z \lesssim 15$ requires a more sophisticated galaxy formation model, which in turn introduces a significantly larger number of free parameters~\cite{chakrabortyModellingStarformationActivity2024, choudhuryCapturingSmallScaleReionization2025}. This added complexity renders an MCMC-based parameter inference using high-resolution simulations computationally infeasible, preventing a fair comparison with our ANN-based results. In future, we aim to extend our framework to a fourteen-parameter model~\cite{choudhuryCapturingSmallScaleReionization2025}, incorporating the latest UVLFs.
    
    \item We include estimates of the thermal state of the low-density IGM. These are parameterized using the temperature-density relation, which for low-density gas can be well approximated by a power-law relation
    \begin{equation}
        T=T_0\Delta^{\gamma - 1},
    \end{equation}
    where $\Delta$ is the overdensity, $T_0$ is the temperature at the mean density, and $\gamma$ is the slope of the relation \cite{huiEquationStatePhotoionized1997}. We use observational estimates of $T_0$ and $\gamma$ at redshifts $z=5.4,\,5.6,\text{ and }5.8$, inferred from the spike statistics of the Ly$\alpha$ transmitted flux~\cite{gaikwadProbingThermalState2020}.
    
    \item Finally, we exclude models that produce reionization histories exhibiting ``double reionization'', in which the ionized fraction undergoes a significant decrease after an initial rise before reaching unity again. Such scenarios are incompatible with our temperature-evolution model, which assumes photoionization equilibrium; this assumption breaks down if regions recombine more rapidly than they are reionized. In our parametrization, these histories arise for extreme negative values of $\alpha$, which cause the ionization efficiency $\zeta(z)$ to be very high at high redshifts and to decline sharply toward lower redshifts. In combination with a moderate value of $\zeta_0$, this can lead to an early phase of efficient ionization, followed by a decrease in the ionization fraction due to recombinations as $\zeta(z)$ decreases.
    
\end{enumerate}

\subsection{Parameter constraints using conventional MCMC} \label{sec:conventional_constraints}

We use an MCMC-based approach to constrain our model's free parameters which involves estimating the posterior $\mathcal{P}(\theta|\mathcal{D})$, which is the conditional probability distribution of the model parameters $\theta$, given an observational dataset $\mathcal{D}$. This is given by the Bayes' theorem
\begin{equation}
    \mathcal{P}(\theta|\mathcal{D}) = \frac{\mathcal{L}(\mathcal{D}|\theta)\pi(\theta)}{\mathcal{P}(\mathcal{D})},    
\end{equation}
where $\mathcal{L}(\mathcal{D}|\theta)$ is the likelihood (i.e., the conditional probability distribution of data given the parameters), $\pi(\theta)$ is the prior, and $\mathcal{P}(\mathcal{D})$ is the evidence.

For all observational constraints, except (\ref{constraint:darkpixels}) and (\ref{constraint:reionzend}), we use a multidimensional Gaussian likelihood of the form
\begin{equation}
    \mathcal{L}(\mathcal{D}|\theta) = \prod_i \exp\left[-\frac{1}{2}\left(\frac{\mathcal{D}_i-\mathcal{M}_i(\theta)}{\sigma_i}\right)^2\right],
\end{equation}
where the index $i$ runs over all the data points used. For data with asymmetric error bars, we use the upper uncertainty when the model estimate lies above the data point, and the lower uncertainty when it lies below. For constraints (\ref{constraint:darkpixels}) and (\ref{constraint:reionzend}), where we have limits instead of measurements, we use a likelihood given by~\cite{gharaConstrainingIntergalacticMedium2020,maityConstrainingReionizationThermal2022}
\begin{equation}
    \mathcal{L}(\mathcal{D}|\theta)=\prod_i\frac{1}{2}\left[1+\text{erf}\left(\pm \frac{\mathcal{M}_i(\theta)-\mathcal{D}_i}{\sqrt{2}\sigma_i}\right)\right],
\end{equation}
where the $\pm$ sign inside the error function represents the lower and upper limits, respectively. We adopt broad uniform priors for all free parameters, as specified in table~\ref{tab:priors}.

\begin{table}[htbp]
\centering
\begin{tabular}{ll}
\hline
Parameter & Prior Range \\
\hline
$\log \zeta_0$ & [0, 10] \\
$\alpha$ & [$-$20, 20] \\
$\log T_{\mathrm{re}}$ & [2, 5] \\
$\log f_{\mathrm{esc},0}$ & [$-$5, 0] \\
$\beta$ & [$-$1, 0] \\
\hline
\end{tabular}
\caption{Uniform prior ranges for the free parameters.\label{tab:priors}}
\end{table}

To sample the parameter space and generate the posterior probability distributions, we use the publicly available package \texttt{Cobaya}~\cite{torradoCobayaBayesianAnalysis2019a,torradoCobayaCodeBayesian2021a}. Specifically, we use its implementation of the Metropolis-Hastings sampler~\cite{metropolisEquationStateCalculations1953, lewisEfficientSamplingFast2013a}. The samples are drawn using 8 parallel chains~\cite{lewisCosmologicalParametersCMB2002a, lewisEfficientSamplingFast2013a}. The chains are assumed to be converged when the Gelman-Rubin $R-1$ statistic~\cite{gelmanInferenceIterativeSimulation1992} falls below 0.01 for two consecutive evaluations. Once converged, we discard the first 30\% of the samples as burn-in and use the rest for further analysis. The resultant posterior distributions are analyzed using the \texttt{GetDist}~\cite{lewisGetDistPythonPackage2019} package. We use simulations with two different resolutions to constrain the free parameters, one with a grid size of $\Delta x=8~h^{-1}$~cMpc ($32^3$ grid cells), and another with a grid size of $\Delta x=4~h^{-1}$~cMpc ($64^3$ grid cells). The resultant parameter constraints for the two resolutions are shown in red in figures~\ref{fig:constraints_32} and \ref{fig:constraints_64}. The significant computational cost of this full MCMC run, which is quantified later (see table~\ref{tab:resources}), provides the primary motivation for developing the efficient ANN-based emulator framework described in the next section.

\section{ANN-based emulator} \label{sec:ann_emulator}

\begin{figure}[htbp] 
\centering
\includegraphics[width=\textwidth]{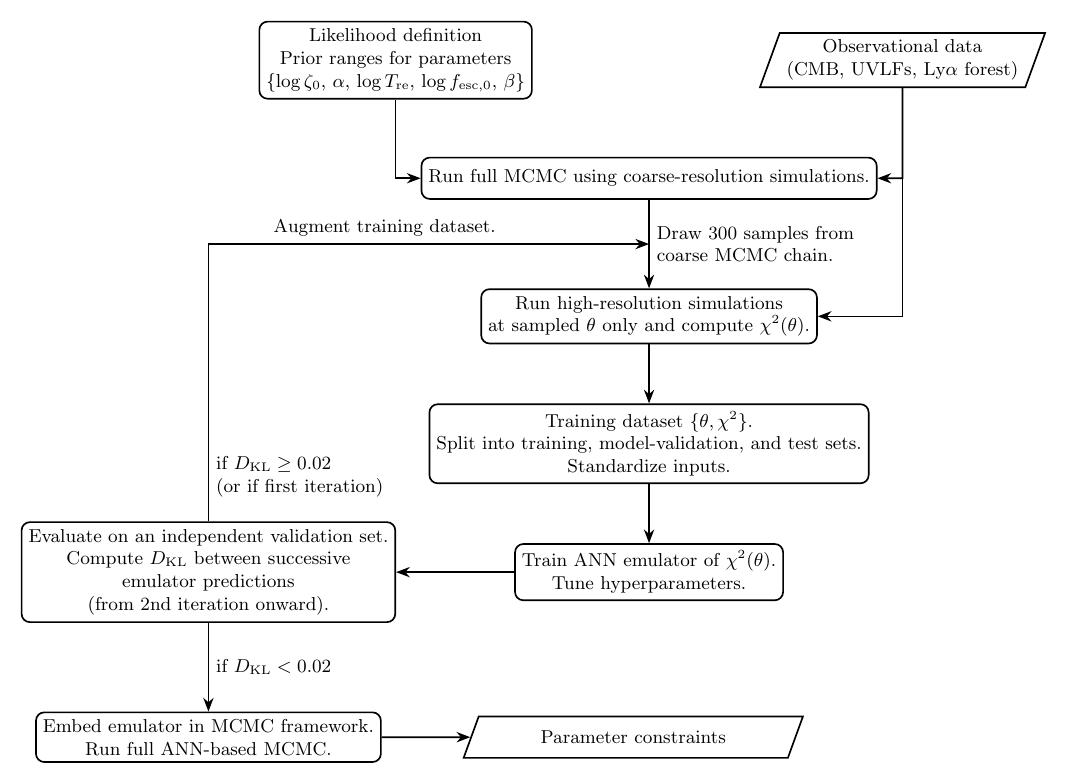}
\caption{Schematic overview of the emulator-based inference framework's methodology and workflow.} \label{fig:overview}
\end{figure}

In this section, we present our ANN-based emulator approach, beginning with a brief overview of ANNs, followed by a detailed description of how they are used to construct emulators of $\chi^2(\theta)$, including dataset generation, training methodology, and integration within an MCMC framework. The same observational constraints and priors used in the previous section are adopted to enable a direct comparison between the two approaches. All the emulators discussed here are implemented using the \texttt{TensorFlow} \texttt{Keras} API~\cite{tensorflow2015-whitepaper,chollet2015keras}. Figure~\ref{fig:overview} provides a schematic overview of the emulator-based inference framework's methodology and workflow.

\subsection{A brief on ANNs} \label{sec:ann_brief}

ANNs are a class of machine learning models that have emerged as a powerful tool for approximating complex, non-linear functions. Inspired by the structure and function of biological neural networks, ANNs consist of layers of interconnected nodes, also called neurons. For the $j^\text{th}$ neuron in the $l^\text{th}$ layer, its output is a weighted sum of its inputs $a_i^{l-1}$ from the previous layer, passed through a non-linear activation function
\begin{equation}
    a_j^{l}=h\left(\sum_{i=1}w_{ji}a_i^{l-1}+b_j\right),
\end{equation}
where $w_{ji}$ are the weights, $b_j$ are the biases, and $h$ is a non-linear activation function. A sufficiently large and well-trained ANN can approximate any continuous function $f(x)$ in a given domain. To train an ANN as an emulator, one requires a training dataset consisting of input-output pairs $\{x, y\}$ that effectively represent the mapping $f(x)=y$ in the domain of interest. Training involves minimizing a loss function such as the mean squared error (MSE)
\begin{equation}
    \text{MSE} = \frac{1}{n}\sum_{i=1}^n\left(y_{i,\mathrm{true}}-y_{i,\mathrm{predicted}}\right)^2,
\end{equation}
by iteratively updating the weights and biases of the network over multiple training cycles, called epochs, with updates occurring after processing mini-batches of data. The training uses an algorithm called backpropagation~\cite{rumelhartLearningRepresentationsBackpropagating1986}, which applies the chain rule to compute gradients of the loss function with respect to the weights and biases. These gradients are then used by an optimization algorithm to update the network's weights and biases. The step size of these updates is controlled by the learning rate, which determines how quickly the model adjusts its weights during training.

The number of layers, number of neurons per layer, and learning rate are examples of ``hyperparameters'' of an ANN that must be chosen before training. Other important hyperparameters include the choice of activation functions, batch size, and the optimization algorithm. Choosing the right hyperparameters is essential to obtaining a well-performing model, as they can significantly impact its accuracy.

\subsection{General methodology for emulator training and MCMC integration}
\label{sec:training}

Building upon the basic principles outlined above, our primary objective is to train an ANN-based emulator that can accurately and rapidly interpolate the chi-squared value, $\chi^2(\theta)$, for any given parameter vector $\theta$ within the prior volume. In particular, the emulator must interpolate accurately in the high-likelihood (low $\chi^2$) region, where precise evaluations of $\chi^2(\theta)$ are most critical for parameter inference. This trained emulator can then be embedded within a conventional MCMC framework to efficiently explore the parameter space. This section outlines the general framework we use for training, validating, and deploying these emulators.

We choose to emulate $\chi^2(\theta)$ directly rather than the individual observables entering the likelihood. This approach significantly simplifies the learning task: $\chi^2$ is expected to be a smooth scalar function of the parameters, avoiding the challenges of jointly emulating several observables with different dynamic ranges and noise properties. This choice has also proven robust in earlier emulator frameworks for \texttt{SCRIPT}~\cite{maityFastMethodReionization2023,choudhuryGPRbasedEmulatorSeminumerical2024}, and it provides a scalable strategy for higher-dimensional extensions of the model. However, emulating the likelihood directly also has some limitations compared to emulating individual observables. In particular, the emulator is tied to a fixed observational dataset and likelihood definition, and would need to be retrained if either were modified. In addition, errors in the emulator predictions are not easily attributable to specific observables, reducing interpretability compared to observable-level emulators. Nevertheless, for the present work, these limitations are outweighed by the practical advantages of emulating the scalar $\chi^2(\theta)$. In particular, this choice allows us to use a considerably simpler ANN architecture and, consequently, achieve significantly faster and more stable training.

The emulator is implemented as a multilayer perceptron (MLP) with Rectified Linear Unit (ReLU) activations in all hidden layers. Optimization is performed using the AdamW optimizer~\cite{loshchilovDecoupledWeightDecay2019}, with the weight decay factor fixed at its default value of 0.004, and all layer weights initialized using the Glorot scheme~\cite{glorotUnderstandingDifficultyTraining2010}. A given training dataset (described in section \ref{sec:constructing}) is partitioned into training, model-validation, and test subsets in the ratio 6:2:2. The input parameter vectors are then standardized by subtracting, for each parameter, its mean value over the training set and rescaling by the corresponding standard deviation. These scaling parameters are computed using only the training set and then applied unchanged to the model-validation and test sets, ensuring that no information from the evaluation data is used during training. The batch size is chosen to scale with the size of the training set and is set to the integer part of its square root.

During each training stage, a grid of candidate networks is trained on the training set, systematically varying the number of layers, the number of neurons per layer, and the learning rate, with the explored ranges listed in table~\ref{tab:hyperparams}. To mitigate the effect of unfavorable weight initialization, which can sometimes yield poor results even for otherwise well-chosen hyperparameters, each candidate network is trained twice using different random seeds. Training continues until the MSE on the model-validation set fails to decrease for 50 consecutive epochs, implementing early stopping to prevent overfitting. Once all candidate networks have been trained, they are evaluated on the test set, and the best-performing emulator is chosen as the trained network with the largest value of
\begin{equation}
    R^2 = 1- \frac{\sum_{\{\theta\}_\mathrm{test}}(\chi^2_{\mathrm{true}}(\theta)-\chi^2_{\mathrm{predicted}}(\theta))^2}{\sum_{\{\theta\}_\mathrm{test}}(\chi^2_{\mathrm{true}}(\theta)-\bar\chi^2_{\mathrm{true}})^2},
\end{equation}
where $\chi^2_\mathrm{true}$ and $\chi^2_\mathrm{predicted}$ are the true and predicted $\chi^2$ values, and $\bar\chi^2_{\mathrm{true}}$ is the mean of the true $\chi^2$ values in the test set.

\begin{table}[htbp]
\centering
\begin{tabular}{ll}
\hline
Hyperparameter & Values \\
\hline
Number of layers & 1, 2, 3 \\
Number of neurons per layer & 100, 200, 300, 400, 500 \\
Learning rate & $10^{-4},\,10^{-3}$ \\
\hline
\end{tabular}
\caption{Range of hyperparameter values for which candidate networks were trained.\label{tab:hyperparams}}
\end{table}

The final best-performing emulator is then embedded within a conventional MCMC framework to generate the posterior probability distributions using the same \texttt{Cobaya} configuration as described in section \ref{sec:conventional_constraints}. For all ANN-based MCMC runs, we use the same broad, uniform priors specified in table~\ref{tab:priors}. This complete methodology provides a consistent basis for training an emulator and rigorously testing its performance against the benchmark conventional MCMC.

\subsection{Training dataset construction and performance} \label{sec:constructing}

Generating the training dataset is the most critical step in emulator construction, as it directly determines the accuracy and efficiency of subsequent parameter inference. To enable the emulator to interpolate $\chi^2(\theta)$ for any parameter vector $\theta$, we require a representative training dataset of $\{\theta,\,\chi^2\}$ pairs, whose quality depends critically on two choices: where in parameter space the vectors are sampled from, and how many such samples are included. The parameter vectors selected for training must ensure dense coverage of the high-likelihood regions, while avoiding the computational cost of evaluating points that contribute little to improving emulator accuracy.

Because our priors are broad, conventional space-filling designs such as random or LH sampling are inefficient: they distribute points uniformly over the prior volume and thus sample the high-likelihood region sparsely. For reference, we include our results using the LH sampling approach in section~\ref{sec:latin_results}. 

\subsubsection{Results using Latin Hypercube sampling} \label{sec:latin_results}

\begin{figure}[htbp] 
\centering
\includegraphics[width=\textwidth]{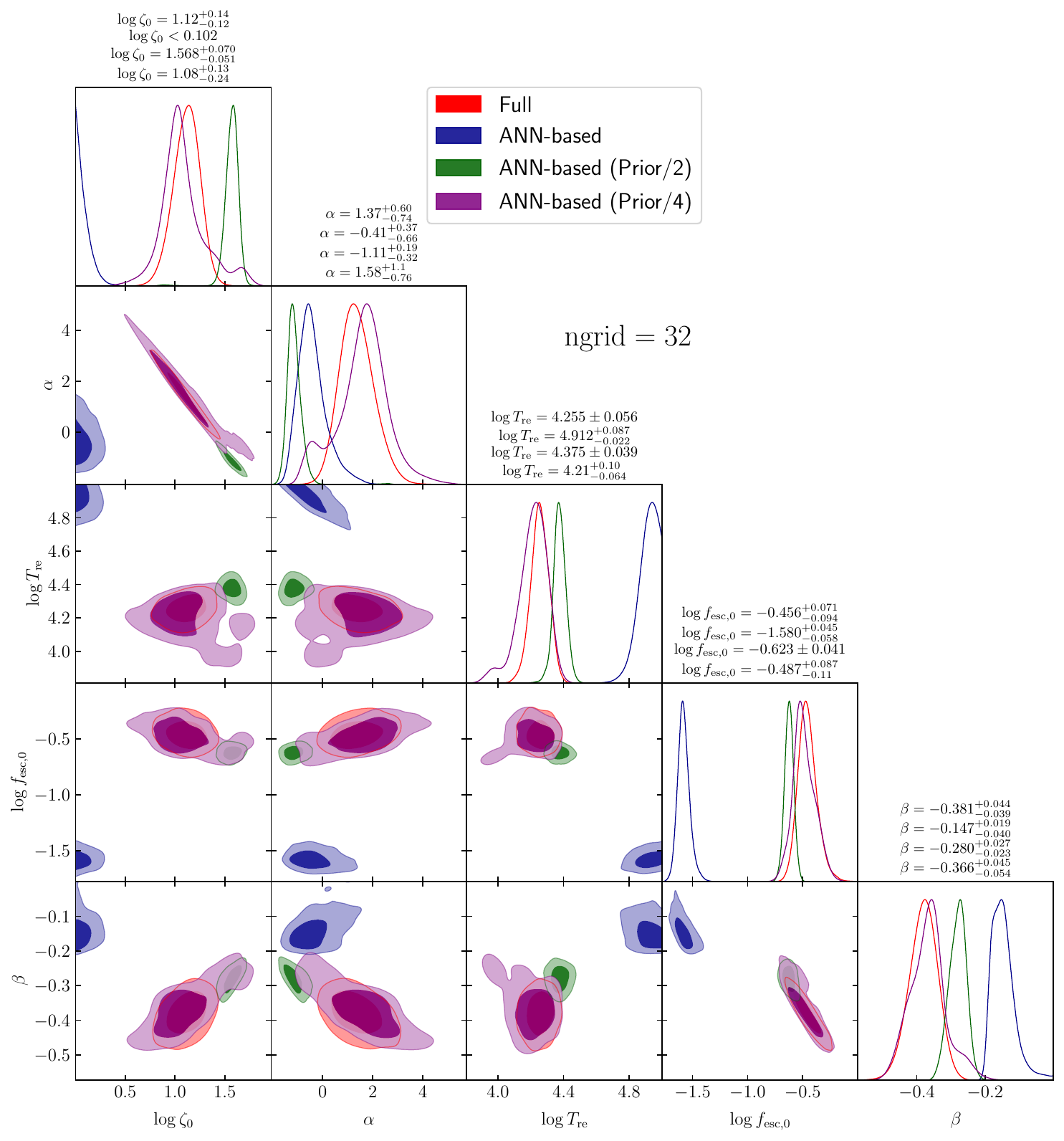}
\caption{Parameter constraints from the full high-resolution ($N_\text{grid}=32$) MCMC run (red) and the ANN-based MCMC runs with the emulator trained on $10^4$ parameter vectors sampled using LH sampling from the full prior, and with the priors reduced by factors of 2 and 4.} \label{fig:constraints_latin_32}
\end{figure}

Here we investigate a strategy for building the training dataset, where the parameter vectors $\theta$ are sampled directly from the uniform prior $\pi(\theta)$ defined over the ranges listed in table~\ref{tab:priors}. Specifically, we randomly sample $10^4$ parameter vectors using LH sampling from the prior volume and evaluate their corresponding $\chi^2$ values by comparing the model predictions (computed with $N_\mathrm{grid}=32$ simulations) to the observational constraints. 

Within this dataset, we find many instances of extremely large $\chi^2$ values, including some cases where $\chi^2$ diverges to infinity. These occur when the sampled parameters produce model outputs that violate one or more observational constraints, which is an expected consequence of the prior ranges being much broader than the spread of the posterior. However, this wide dynamic range of $\chi^2$ values poses challenges for training ANNs effectively. To mitigate this, we discard the samples with infinite $\chi^2$ and use $\log \chi^2$ values instead for the remaining samples, which compresses the range and improves numerical stability during training. At the inference stage, the emulator predicts $\log \chi^2$ values, which we exponentiate to recover the corresponding $\chi^2$ needed for likelihood evaluation. The final training dataset of $\{\theta,\,\log \chi^2\}$ pairs is used to train an emulator using the same procedure as described in section \ref{sec:training}. 

Additionally, we repeat this process using another two sets of $10^4$ parameter vectors sampled using LH sampling, but with the priors artificially reduced by factors of 2 and 4. This exercise is purely illustrative, as such prior reductions are generally not feasible when the parameters are highly unconstrained. The resultant parameter constraints are shown in figure \ref{fig:constraints_latin_32}. It is clearly seen that the emulators do not perform well when trained using LH samples from broad priors, and can lead to unreliable inference results resulting from spurious posterior distributions. While this exercise highlights the limitations of uniform space-filling designs, it also motivates a more efficient approach that concentrates sampling within the regions of high-likelihood.

The failure demonstrated in figure~\ref{fig:constraints_latin_32} is not necessarily a failure of the ANN itself, but a failure of the training data. It clearly illustrates that for broad priors, a space-filling design like LH sampling cannot provide adequate coverage of the high-likelihood region without an intractably large number of training points. This result directly motivates our two-part solution in the following section, which is designed to first locate this high-likelihood region efficiently, and second, adaptively sample it to build a predictive emulator.

\subsubsection{Training set using targeted sampling}

In this work, we instead follow the targeted sampling strategy introduced in~\cite{maityFastMethodReionization2023}, which takes advantage of two features of our setup: (i) the large-scale properties of our model are convergent with respect to simulation resolution, and (ii) full MCMC runs with coarse-resolution simulations are computationally inexpensive. This allows us to sample parameter vectors for the training dataset directly from the MCMC chain of a full MCMC run using inexpensive coarse-resolution simulations, ensuring that samples are concentrated in the high-likelihood regions of parameter space.

The second consideration is to determine an appropriate training dataset size without oversampling. We address this using an adaptive procedure. We begin by randomly selecting 300 parameter vectors from the coarse-resolution MCMC chain and computing their $\chi^2$ values by comparing the model predictions (computed with simulations run at the target high-resolution) to the observational constraints. This initial dataset is used to train an emulator (the training procedure is described in section~\ref{sec:training}). 

In addition, we define a fixed validation set, $\{\theta\}_\mathrm{val}$, at the outset. This set is used exclusively to evaluate the convergence of the adaptive sampling procedure and is distinct from the ``model-validation'' set described in section \ref{sec:training}, which is used during network training for early stopping. This set is constructed by drawing samples with probability proportional to their weights, ensuring that it provides an unweighted sample representative of the full posterior distribution. Crucially, this set is never used during training, as doing so would cause data leakage, where information from the validation set inadvertently influences training, resulting in overly optimistic performance estimates and poor generalization to unseen parameters. After each training stage, the emulator predicts $\chi^2$ for the validation set. The training dataset is then augmented with 300 additional samples, a new emulator is retrained, and the validation predictions are updated. To measure convergence, we compute the KL divergence between successive validation predictions,
\begin{equation}
D_\mathrm{KL}(P_{i}||P_{i-1}) = \sum_{\{\theta\}_\mathrm{val}} P_{i}(\theta) \log\frac{P_{i}(\theta)}{P_{i-1}(\theta)},
\end{equation}
where $P_i$ and $P_{i-1}$ are the normalized $\exp(-\chi^2/2)$ predictions from the current and previous iterations, respectively:
\begin{equation}
P_{i}(\theta) = \frac{\exp(-\chi^2_i(\theta)/2)}{\sum_{\{\theta\}_\mathrm{val}} \exp(-\chi^2_i(\theta)/2)}, \quad
P_{i-1}(\theta) = \frac{\exp(-\chi^2_{i-1}(\theta)/2)}{\sum_{\{\theta\}_\mathrm{val}} \exp(-\chi^2_{i-1}(\theta)/2)}.
\end{equation}
The process is repeated until $D_\mathrm{KL}$ falls below 0.02, indicating that the emulator's predictions have stabilized and further sampling would provide negligible benefit. 

We verified that the convergence threshold used in our adaptive procedure ($D_\mathrm{KL} < 0.02$) is robust by varying this threshold within a reasonable range (0.01--0.05). Lowering the threshold naturally increases the size of the final training dataset, with a difference of $\sim 10^3$ samples between threshold values 0.05 and 0.01. However, despite this variation in training dataset size, the recovered posterior remains effectively unchanged across the full range of thresholds. Likewise, increasing the training set beyond the adaptively chosen size yields negligible improvement in $R^2$. Based on these tests, we adopt $D_\mathrm{KL} < 0.02$ as a conservative and efficient stopping criterion.

\begin{figure}[htbp]
    \centering
    \begin{subfigure}{0.49\textwidth}
        \centering
        \includegraphics[width=\textwidth]{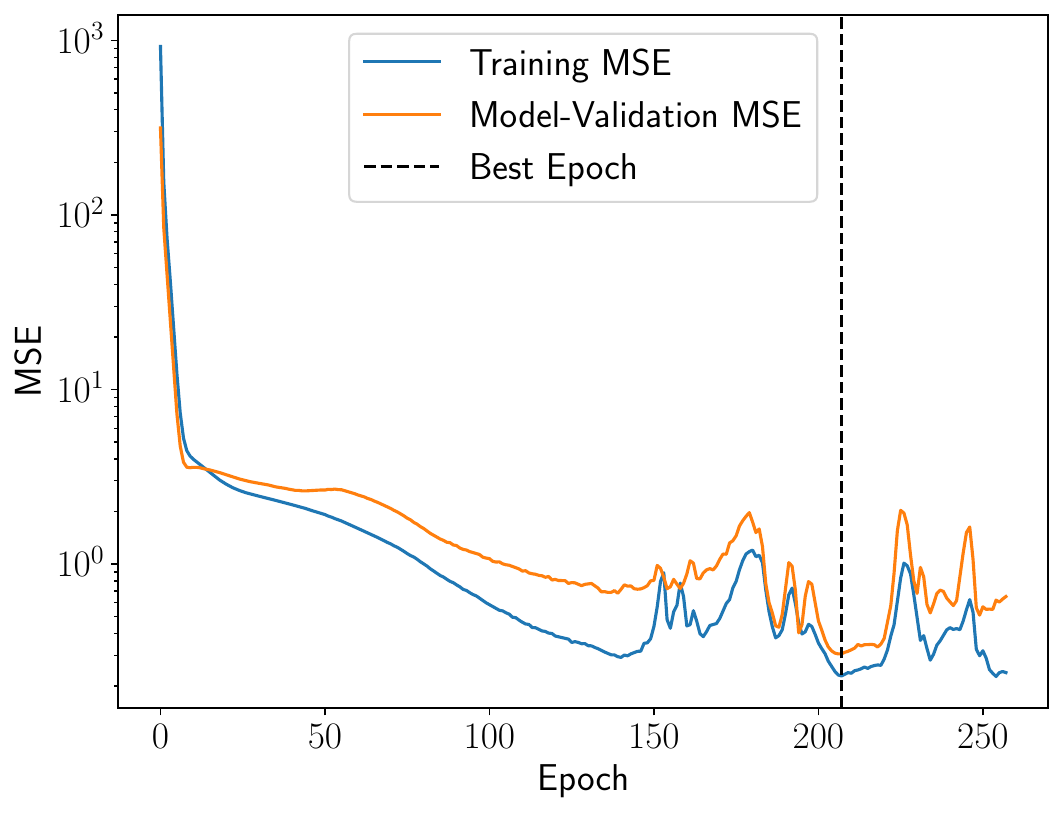}
    \end{subfigure}
    \hfill
    \begin{subfigure}{0.49\textwidth}
        \centering
        \includegraphics[width=\textwidth]{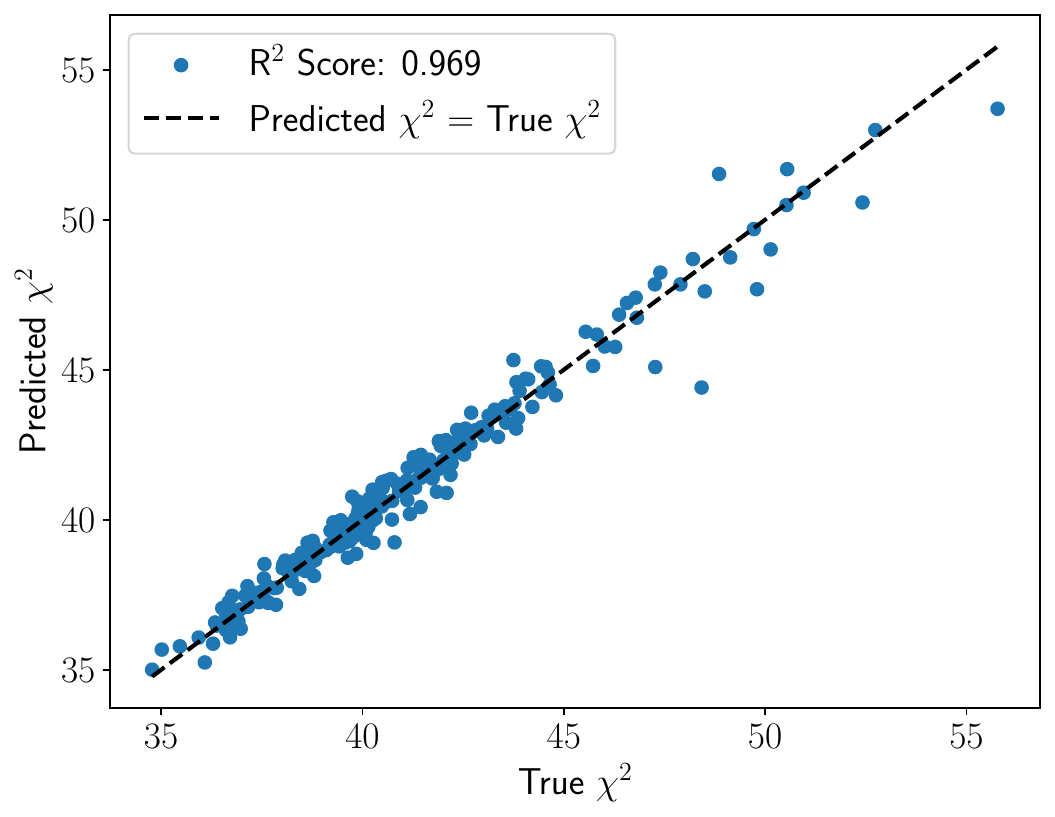}
    \end{subfigure}
    \caption{Training diagnostic plots for the best-performing $N_\text{grid}=32$ emulator. \textit{Left}: Evolution of the training and model-validation MSE as a function of training epochs. The dashed line marks the epoch at which the model-validation MSE reached its minimum, at which point the trained network was saved. \textit{Right}: True $\chi^2$ values versus the $\chi^2$ values predicted by the emulator for the test set.}
    \label{fig:model_perf_32}
\end{figure}

\begin{figure}[htbp]
    \centering
    \begin{subfigure}{0.49\textwidth}
        \centering
        \includegraphics[width=\textwidth]{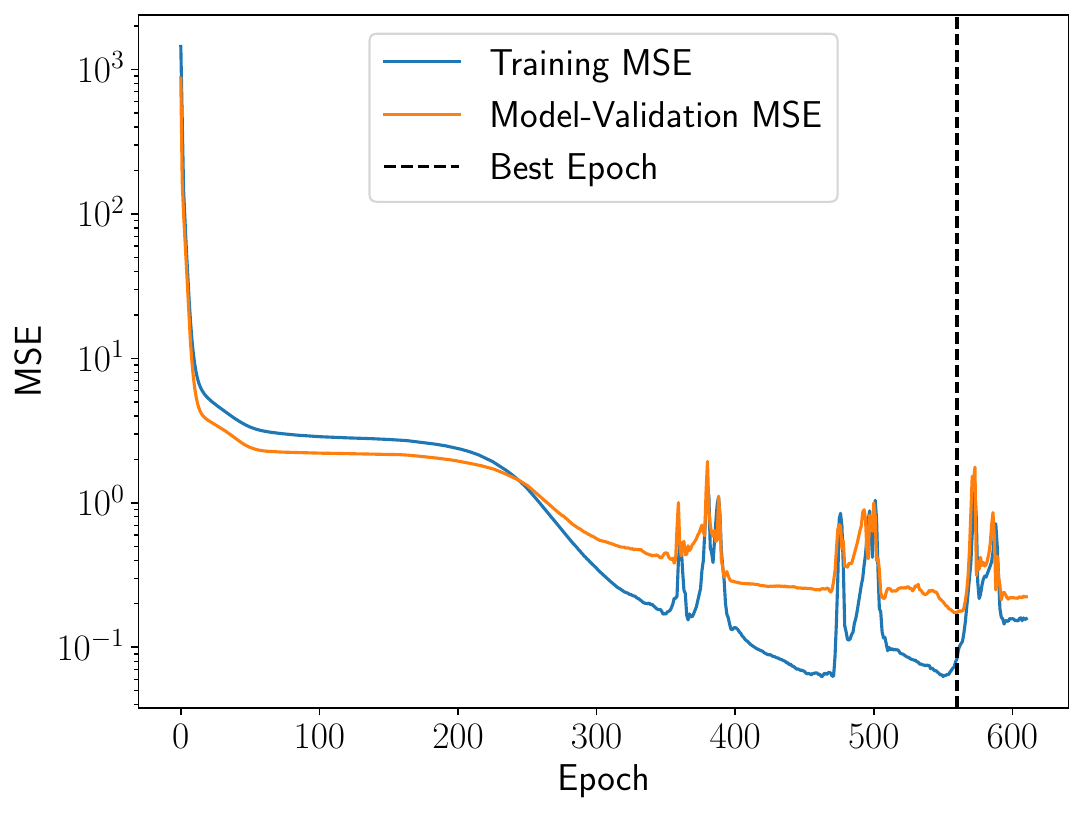}
    \end{subfigure}
    \hfill
    \begin{subfigure}{0.49\textwidth}
        \centering
        \includegraphics[width=\textwidth]{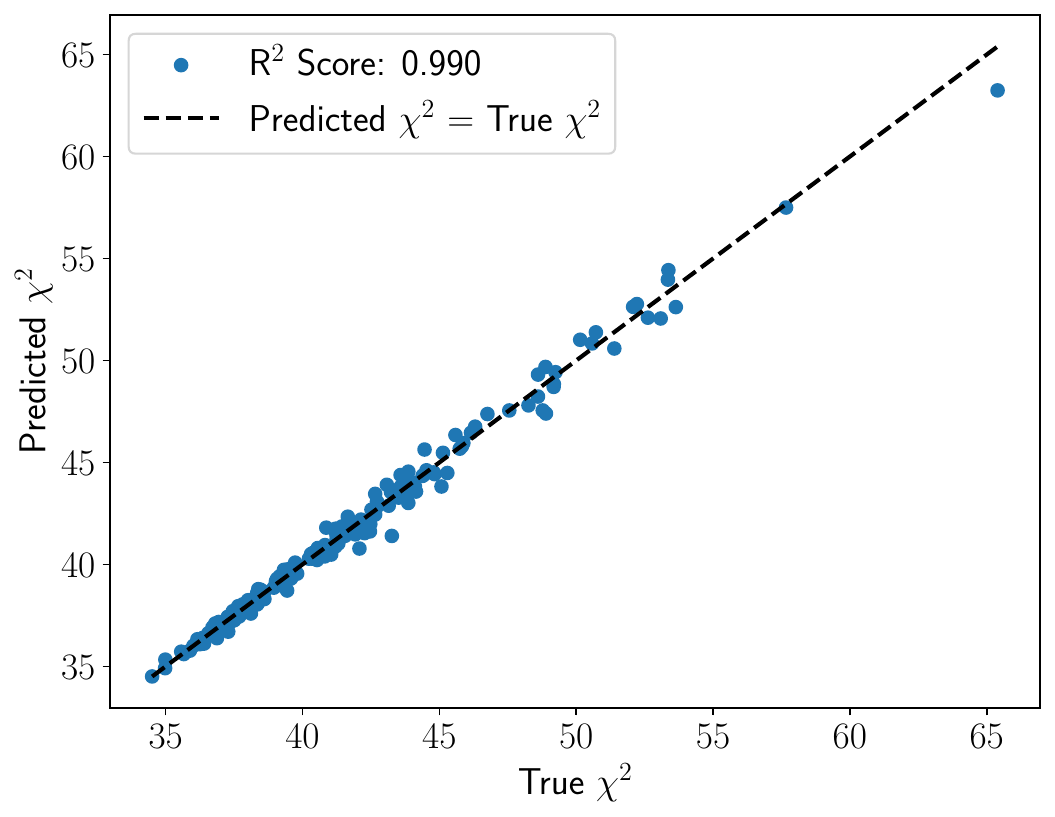}
    \end{subfigure}
    \caption{Training diagnostic plots for the best-performing $N_\text{grid}=64$ emulator. Formatting is the same as in figure \ref{fig:model_perf_32}.}
    \label{fig:model_perf_64}
\end{figure}

\subsection{Parameter constraints using ANN-based MCMC} \label{sec:constraints_ann}

\begin{figure}[htbp] 
\centering
\includegraphics[width=\textwidth]{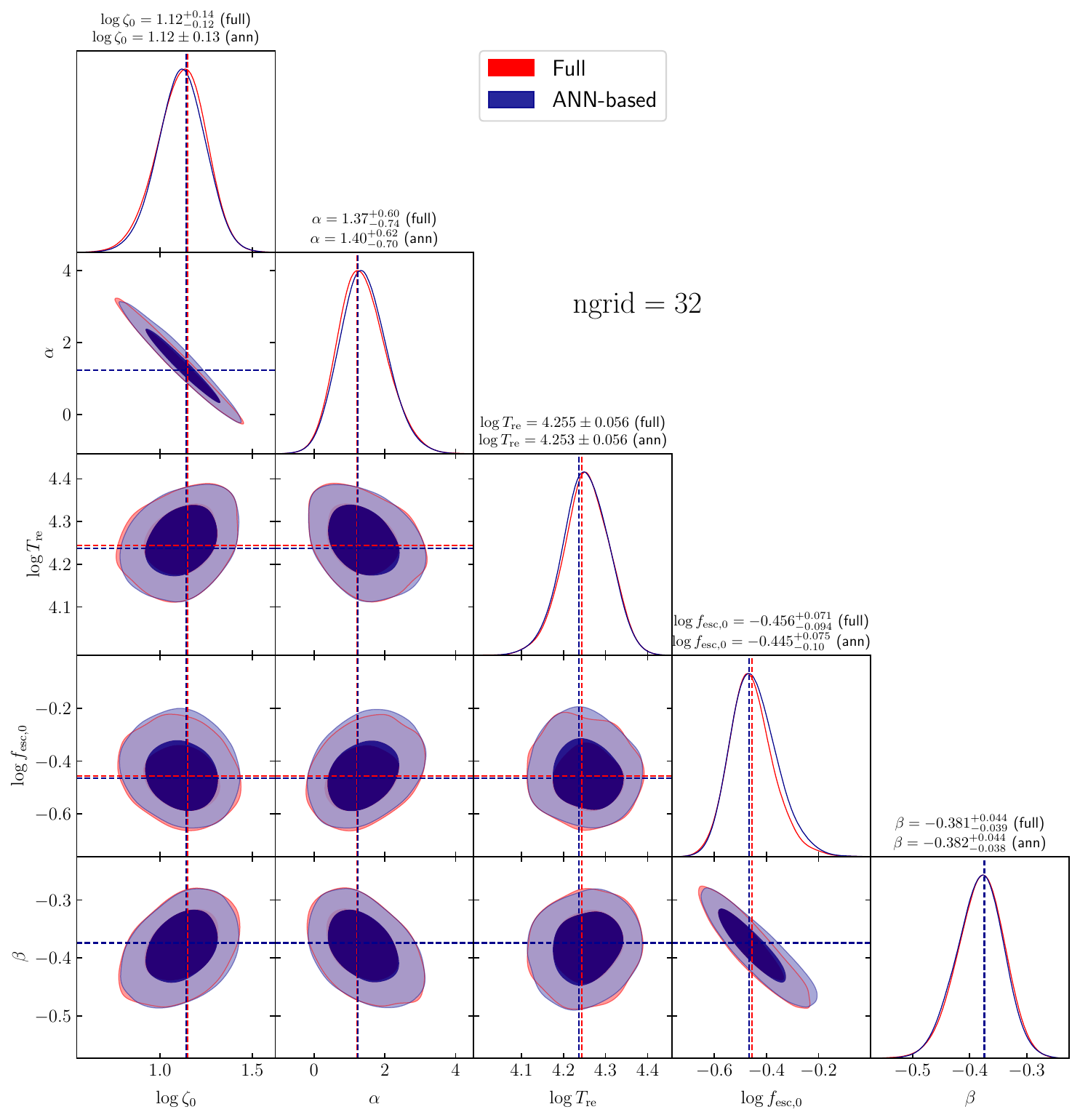}
\caption{Parameter constraints from the full high-resolution ($N_\text{grid}=32$) MCMC run (red) and the ANN-based MCMC run (blue). The diagonal panels show 1D posterior probability distributions while the off-diagonal panels show joint 2D posteriors. The contours represent 68 percent and 95 percent confidence intervals. The dashed lines denote the best-fitting parameter values. The quoted values on each parameter show the mean along with the 1$\sigma$ uncertainties.} \label{fig:constraints_32}
\end{figure}

\begin{figure}[htbp] 
\centering
\includegraphics[width=\textwidth]{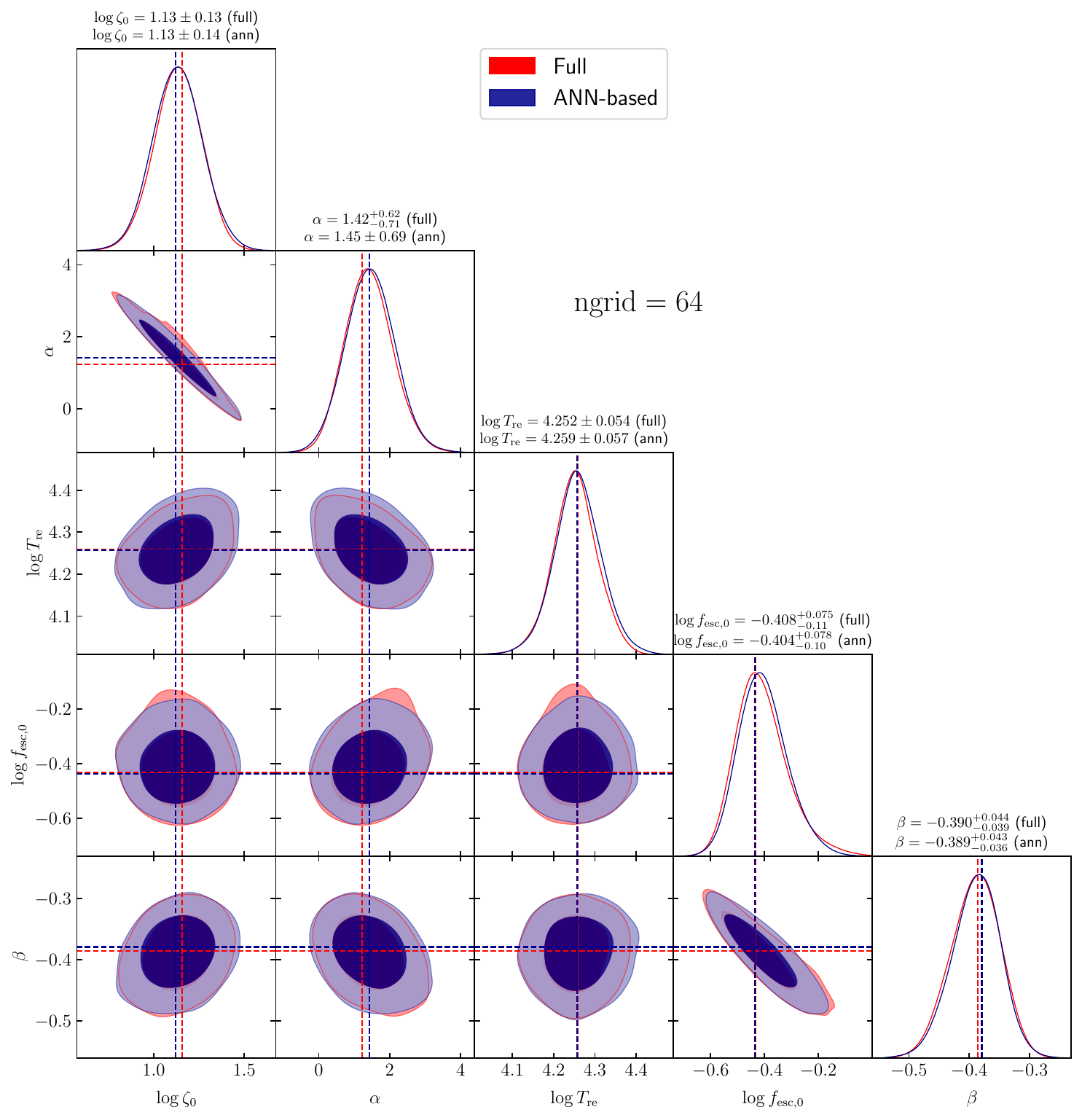}
\caption{Parameter constraints from the full high-resolution ($N_\text{grid}=64$) MCMC run (red) and the ANN-based MCMC run (blue). Formatting is the
same as in figure~\ref{fig:constraints_32}.} \label{fig:constraints_64}
\end{figure}

Having constructed and validated the emulator, we now integrate it within an MCMC framework to perform full parameter inference. This allows us to directly compare the emulator-based posterior distributions with those obtained from the conventional high-resolution MCMC analysis presented in section~\ref{sec:conventional_mcmc}.

For our setup, convergence of the adaptive sampling procedure required a training dataset size of 1200 samples for the $N_\text{grid}=32$ emulator, and 900 for the $N_\text{grid}=64$ emulator. The optimal hyperparameters are summarized in table~\ref{tab:hyperparams_final}. These hyperparameters were selected for optimization because they dominantly control the model's expressive capacity and the convergence properties of the optimizer. Other hyperparameters (e.g., weight decay and batch size) were held fixed to limit the computational cost of the search and to avoid introducing degeneracies, particularly between batch size and learning rate. In addition, since early stopping already provides effective regularization, we fix the weight decay at its default value rather than tuning it explicitly. The best-performing emulators achieved $R^2$ scores of 0.969 and 0.990 on the test set, for $N_\text{grid}=32$ and 64, respectively. Additional diagnostics for the best-performing emulators, including the evolution of training and model-validation MSE and a scatter plot comparing true versus predicted $\chi^2$ values on the test set, are presented in figures \ref{fig:model_perf_32} and \ref{fig:model_perf_64}.

\begin{table}[htbp]
\centering
\begin{tabular}{c|ccc}
\hline
$N_\text{grid}$ & Number of layers & Number of neurons & Learning rate \\
\hline
32 & 3 & [500, 400, 100] & $10^{-3}$ \\
64 & 2 & [500, 200] & $10^{-3}$ \\
\hline
\end{tabular}
\caption{Hyperparameters of the best-performing emulators.\label{tab:hyperparams_final}}
\end{table}

We incorporated the best-performing emulators described above into a conventional MCMC framework and used this new ANN-based framework to repeat the analyses described in section \ref{sec:conventional_constraints} for both $N_\text{grid}=32$ and 64. In figure \ref{fig:constraints_32} and \ref{fig:constraints_64}, we show the comparison of the posteriors obtained using the conventional full MCMC at high-resolution (red) and the ANN-based MCMC (blue). The same priors were used for both cases. The figure demonstrates that the posterior distributions of the parameters obtained from both methods are in close agreement, establishing the reliability of the ANN-based MCMC framework.

A quantitative comparison of the computational requirements of the two approaches is presented in table~\ref{tab:resources}. The table lists the total number of high-resolution simulations and the corresponding CPU hours required by each method. For the ANN-based MCMC, these figures account for the initial coarse-resolution MCMC, generation of the training dataset, emulator training, hyperparameter search, and the final ANN-based MCMC run. Compared to the conventional full MCMC, the ANN-based MCMC run requires $\sim100$ times less number of high-resolution simulations runs. In terms of total CPU hours, the savings correspond to a factor of $\sim5$ for $N_\text{grid}=32$ and a factor of $\sim70$ for $N_\text{grid}=64$. Consequently, the ANN-based framework achieves the same level of accuracy as the conventional method at a dramatically reduced computational cost, thereby providing a far more efficient and scalable approach for exploring large parameter spaces. 

The results in table~\ref{tab:resources} constitute the primary validation of this framework. For the $N_\text{grid} = 64$ case, the ANN-based approach yields approximately a $\sim 70$-fold reduction in total CPU hours while producing posterior constraints that are statistically indistinguishable from those obtained with the full forward model. This improvement is not merely incremental; it demonstrates that the method is capable of transforming the computational landscape of reionization inference. In particular, it establishes the feasibility of analyses that would otherwise be prohibitive, such as constraining the fourteen-parameter model~\cite{choudhuryCapturingSmallScaleReionization2025} required for JWST datasets, which remains computationally intractable using conventional MCMC techniques.

\begin{table}[htbp]
\centering
\begin{tabular}{c|cc|cc}
\hline
\multirow{2}{*}{$N_\text{grid}$} & 
\multicolumn{2}{c|}{Number of (high-res) simulations} & 
\multicolumn{2}{c}{Total CPU hours} \\
& Full & ANN-based & Full & ANN-based \\
\hline
32 & 80364 & 1200 & 374 & 82 \\
64 & 114686 & 900 & 8184 & 112 \\
\hline
\end{tabular}
\caption{Comparison of computational requirements for Full and ANN-based MCMC.\label{tab:resources}}
\end{table}

\section{Summary and future outlook} \label{sec:summary}

In this work, we have developed an ANN-based framework for efficiently constraining parameters of the EoR using the semi-numerical, photon-conserving reionization model \texttt{SCRIPT}. The model self-consistently accounts for inhomogeneous recombinations, IGM thermal history, and radiative feedback, and is described by five free parameters that govern the ionizing efficiency, reionization heating, and the escape fraction of ionizing photons.

A key challenge in emulator construction is generating a training dataset that adequately samples the relevant parameter space. Conventional space-filling designs, such as random or LH sampling, are inefficient when prior ranges are broad, since many simulations are wasted in low-likelihood regions. To address this, we introduced a two-step strategy. First, we use full, reliable MCMC using inexpensive coarse-resolution simulations to identify high-likelihood regions efficiently, taking advantage of the resolution convergence of \texttt{SCRIPT}. Second, we employ an adaptive procedure that determines the ideal training dataset size by iteratively enlarging the training dataset and comparing successive emulators until their predictions converge.

This approach enables us to train accurate ANN-based emulators using only $\approx 10^3$ high-resolution simulations, achieving predictive accuracies of $R^2 \approx 0.97$--$0.99$. When embedded within a conventional MCMC framework, the ANN-based approach can reproduce the full posterior distributions while reducing the number of expensive high-resolution simulations by a factor of $\sim100$ and the overall computational cost by factors of up to $\sim70$. Our results show that combining coarse-resolution MCMC with adaptive training dataset construction provides a scalable and efficient alternative to conventional space-filling sampling designs, enabling fast and accurate emulator-based inference even when the priors are broad, and the parameters weakly constrained.

We note that while this work focused on a feed-forward ANN, the field is rapidly adopting more advanced machine learning techniques, such as Simulation-Based Inference (SBI) using normalizing flows or diffusion models~\cite[e.g.,][]{alsingFastLikelihoodfreeCosmology2019,hassanHIFlowGeneratingDiverse2022,zhaoCanDiffusionModel2023}. These methods, however, still face the fundamental challenge of sampling efficiency. They perform optimally when training points are concentrated in or near the high-probability posterior region. In this context, two methodological elements employed in this work are of direct relevance: the use of a reliable coarse-resolution MCMC to efficiently identify the high-likelihood region of parameter space, and the adaptive determination of the training dataset size based on emulator convergence. Together, these elements provide a general-purpose strategy for generating optimal training or proposal sets for these next-generation inference tools, significantly boosting their efficiency and making them more viable for complex, high-dimensional astrophysical problems.

This substantial gain in computational efficiency is not merely incremental, it is enabling. The five-parameter model explored here serves as a robust proof of concept, demonstrating that our emulator framework faithfully reproduces full high-resolution inference at a fraction of the cost. The most immediate and scientifically pressing application is to extend this approach to more complex models that are currently computationally prohibitive. In particular, our next step is to apply this methodology to the fourteen-parameter framework of ref.~\cite{choudhuryCapturingSmallScaleReionization2025}, which is required to incorporate the latest JWST UV luminosity function measurements in a self-consistent manner. Exploring such a high-dimensional space is effectively impossible with conventional MCMC techniques, and the emulator strategy presented here provides a practical and scalable path forward.

Looking further ahead, the validated framework is well positioned to analyze forthcoming datasets from 21\,cm facilities such as HERA and the SKA. These observations will demand rapid and repeated exploration of increasingly rich parameter spaces, making ANN-based emulators a crucial component of next-generation inference pipelines in the 21\,cm era. In this sense, the framework developed here transforms reionization inference from a computational challenge into a tractable, scalable, and future-ready enterprise.

\acknowledgments
The authors acknowledge support of the Department of Atomic Energy, Government of India, under project no. 12-R\&D-TFR-5.02-0700. The authors thank Aseem Paranjape for valuable discussions.

\paragraph{Data Availability Statement.} The data generated and presented in this paper will be made available upon reasonable request to the corresponding author.

\bibliographystyle{JHEP}
\bibliography{biblio.bib}
\end{document}